\documentclass[12pt]{article}%
\usepackage{amsmath,latexsym}
\usepackage{graphicx}
\usepackage{amsmath}
\usepackage{amsfonts}
\usepackage{amssymb}%
\date{}
\begin{document}

\title{ On the stability of thin-shell wormholes in
noncommutative geometry}
   \author{
\large Peter K.F. Kuhfittig \footnote{kuhfitti@msoe.edu}
    \\ \\
 \small Department of Mathematics, Milwaukee School of
Engineering,
Milwaukee, Wisconsin 53202-3109, USA\\
 }

\maketitle

\begin{abstract}\noindent
This paper reexamines a special class of thin-shell
wormholes that are unstable in general relativity
in the framework of noncommutative geometry.  It is
shown that as a consequence of the intrinsic uncertainty
these wormholes are stable to small linearized radial
perturbations.  Several different spacetimes are
considered.

\phantom{a}
\end{abstract}

\section{Introduction}
An important outcome of string theory is the realization
that coordinates may become noncommuting operators on a
$D$-brane [Witten (1996), Seiberg \& Witten (1999)].
The result is a fundamental discretization of spacetime
due to the commutator
[\textbf{x}$^{\mu}$,\,\,\textbf{x}$^{\nu}]=
i\,\theta^{\mu\nu}$, where $\theta^{\mu\nu}$ is an
antisymmetric matrix, in much the same way as the Planck
constant $\hbar$ discretizes phase space [Gruppuso
(2005)].  Moreover, noncommutativity is an intrinsic
property of spacetime and does not depend on particular
features such as curvature.

It was pointed out by Smailagic \& Spalluci (2003) that
noncommutativity replaces point-like structures by
smeared objects and so may eliminate the
divergences that normally appear in general relativity.
An effective way to model the smearing effect is by
the use of the Gaussian distribution of minimal length
$\sqrt{\alpha}$ instead of the Dirac-delta function.
As a result, according to Nicolini, Smailagic \&
Spalluci (2006), the energy density of the static and
spherically symmetric smeared and particle-like
gravitational source has the form
\begin{equation}\label{E:M}
  \rho(r)=\frac{m}{(4\pi\alpha)^{3/2}}
      e^{-r^2/4\alpha},
\end{equation}
i.e., the mass $m$ is diffused throughout the region of
linear dimension $\sqrt{\alpha}$ due to the uncertainty.
Using this gravitational source in the Einstein field
equations, the line element was found to be
\begin{equation}\label{E:line}
 ds^2=-\left(1-\frac{2m^*(r)}{r}\right)dt^2
   +\frac{dr^2}{1-2m^*(r)/r}+r^2(d\theta^2+
       \text{sin}^2\theta\,d\phi^2),
\end{equation}
where
\begin{equation}\label{E:mass1}
   m^*(r)=\frac{2m}{\sqrt{\pi}}\gamma\left(\frac{3}{2},
    \frac{r^2}{4\alpha}\right)=\frac{2m}{\sqrt{\pi}}
      \int^{r^2/4\alpha}_0\sqrt{t}\,e^{-t}\,dt.
\end{equation}
Here
\begin{equation}\label{E:gamma1}
  \gamma\left(\frac{3}{2},\frac{r^2}{4\alpha}\right)=
      \int^{r^2/4\alpha}_0\sqrt{t}\,e^{-t}\,dt
\end{equation}
is the lower incomplete gamma function.  The classical
Schwarzschild mass is recovered in the limit as
$r\sqrt{\alpha}\rightarrow\infty$.  (Recall that the
lower incomplete gamma function starts at the
origin, rises sharply, and then approaches unity
asymptotically.)

Some modification will be required when applying these
ideas to thin-shell wormholes, discussed in the next
section.  For now we need only to note that the throat
is assumed to be a thin shell, a sphere of radius
$r=a_0$.  So instead of a smeared particle, we have a
smeared surface.

The main purpose of this paper is to show that the special
thin-shell wormholes discussed here are stable to small
linearized radial perturbations given a
noncommutative-geometry framework, even though they are
unstable in the framework of classical general relativity
(GR).  For this reason the concentration will be on the
smeared spherical surface of radius $a_0$ rather than
on smeared point-like structures, since the surface is
directly affected by radial perturbations.

\section{Thin-shell wormholes}

A powerful theoretical method for constructing a class of
spherically symmetric wormholes from black-hole spacetimes
was proposed by Visser (1989).  The starting point is a
spherically symmetric line element
\begin{equation}\label{E:line1}
  ds^2=-f(r)\,dt^2+[f(r)]^{-1}dr^2+r^2(d\theta^2+
       \text{sin}^2\theta\,d\phi^2)
\end{equation}
describing a black-hole spacetime.  The construction
begins with two copies of a black hole and removing
from each the four-dimensional region

\[
  \Omega^\pm = \{r\leq a\,|\,a>r_h\},
\]
where $r=r_h$ is the (outer) event horizon.  The topological
identification of the timelike hypersurfaces
\[
  \partial\Omega^\pm =\{r=a\,|\,a>r_h\}
\]
results in a manifold that is geodesically complete and
possesses two asymptotically flat regions connected by a throat.

A dynamic analysis depends on the Lanczos equations and is now
considered standard.  [See, for example, Dias \& Lemos (2010),
Eiroa (2009), Lemos \& Lobo (2008), Lobo \& Crawford (2004),
Poisson \& Visser (1995), Rahaman, Kalam \&
Chakraborty (2006), Rahaman, Rahman, Rakib \& Kuhfittig
(2010).]
\begin{equation*}\label{E:Lanczos}
  S^i_{\phantom{i}j}=-\frac{1}{8\pi}\left([K^i_{\phantom{i}j}]
   -\delta^i_{\phantom{i}j}[K]\right),
\end{equation*}
where $[K_{ij}]=K^{+}_{ij}-K^{-}_{ij}$ and
$[K]$ is the trace of $K^i_{\phantom{i}j}$.  In terms of the
surface energy density $\sigma$ and the surface pressure
$\mathcal{P}$, $S^i_{\phantom{i}j}=\text{diag}(-\sigma,
\mathcal{P}, \mathcal{P})$.  By letting $r=a$ be a function
of time, it is shown by Poisson \& Visser ((1995) that
\begin{equation}\label{E:sigma}
\sigma = - \frac{1}{2\pi a}\sqrt{f(a) + \dot{a}^2}
\end{equation}
and
\begin{equation}\label{E:P}
  \mathcal{P} =  -\frac{1}{2}\sigma + \frac{1}{8\pi}
   \frac{2\ddot{a} + f^\prime(a) }{\sqrt{f(a) + \dot{a}^2}}.
\end{equation}
(The overdot denotes the derivatives with respect to $\tau$.)
Since $\sigma$ is negative on the sphere, we are dealing
with exotic matter.  Moreover, since the radial pressure
$p$ is zero for a thin shell, the weak energy condition
is obviously violated.

\section{Thin-shell wormholes with a phantom-like
    equation of state}
\subsection{In general relativity}
As noted in the Introduction, we are going to be concerned
with a special type of thin-shell wormholes, analyzed by
Kuhfittig (2010).  They are characterized by having a
``phantom-like" equation of state $\mathcal{P}=\omega\sigma$,
$\omega<0$, on the shell, a natural analogue of the
Chaplygin-gas equation of state used by Eiroa (2009).

On the question of stability to linearized radial
perturbations, we assume, as always, that $r=a$ is a
function of time.  It it readily checked that
\begin{equation*}
   \frac {d}{d \tau} (\sigma a^2) +
       \mathcal{P}\frac{d}{d \tau}(a^2)=0
\end{equation*}
which can also be written
\begin{equation}\label{E:conservation}
  \frac{d\sigma}{da}+\frac{2}{a}(\sigma+\mathcal{P})=0.
\end{equation}
For a static configuration of radius $a_0$, we have
$\dot{a}=0$ and $\ddot{a}=0$.  Given the equation of state
$\mathcal{P}=\omega\sigma$, Eq. (\ref{E:conservation}) can
be solved by separation of variables to yield
\begin{equation}\label{E:solution}
  \sigma(a)=\sigma_0\left(\frac{a_0}{a}
  \right)^{2(\omega+1)},\quad\sigma_0=\sigma(a_0).
\end{equation}

Rearranging Eq. (\ref{E:sigma}), we obtain the equation of
motion
\[
   \dot{a}^2+V(a)=0.
\]
Here the potential $V(a)$ is defined as
\begin{equation}\label{E:potential}
  V(a)=f(a)-[2\pi a\sigma(a)]^2.
\end{equation}
Expanding $V(a)$ around $a_0$, we get
\[
 V(a)=V(a_0)+V'(a_0)(a-a_0)+
 \frac{1}{2}V''(a_0)(a-a_0)^2+\mathcal{O}[(a-a_0)^3].
\]
Since we are linearizing around $a=a_0$, we require that
$V(a_0)=0$ and $V'(a_0)=0$.  The configuration is in
stable equilibrium if $V''(a_0)>0$.

\subsection{In noncommutative geometry}
A discussion of thin-shell wormholes in noncommutative
geometry has to take into account the nature of the
thin shell.  The reason is that we are now dealing with
a surface rather than a point-like structure.  Moreover,
we would expect the surface to be smeared as a
consequence of the intrinsic uncertainty.  So,
returning to Eq. (\ref{E:mass1}), observe that if a
particle is located on the sphere $r=a_0$, then its
mass is given by
\begin{equation}\label{E:mass2}
   m\int^{(r-a_0)^2/4\alpha}_0\frac{2}
     {\sqrt{\pi}}\,\sqrt{t}\,e^{-t}\,dt.
\end{equation}
Here
\begin{equation}\label{E:gamma2}
\gamma\left(\frac{3}{2},\frac{(r-a_0)^2}{4\alpha}\right)
   =\int^{(r-a_0)^2/4\alpha}_0\sqrt{t}\,e^{-t}\,dt
\end{equation}
is the corresponding lower incomplete gamma function,
a pure translation of Eq. (\ref{E:gamma1}) by a
distance $r=a_0$ in the $r$-direction, i.e.,
independent of $\theta$ and $\phi$, as shown, for
example, by
Rahaman, Kuhfittig, Chakraborty, Usmani \& Ray (2012).
To visualize the process, one can simply choose a ray
in a particular direction: now the function starts at
$r=a_0$ instead of the origin and approaches $m$
asymptotically along the ray.  The concentration
on the radial direction is appropriate because we are
interested in linearized radial perturbations.

The distance to a smeared object is necessarily
smeared.  Given the nature of the smearing in
noncommutative geometry, we may assume that a
smeared distance is proportional to the lower
incomplete gamma function.  The reason for this
can also be seen from the following heuristic
argument: Consider $m^*(r_0)$ for some arbitrary
fixed $r_0$.  Then the proper distance $\ell$
between two points is from line element
(\ref{E:line})
\begin{equation}\label{E:proper}
  \ell=\int^b_a\frac{dr}{\sqrt{1-2m^*(r_0)/r}}
  =m^*(r_0)\int^b_a\frac{dr}
  {m^*(r_0)\sqrt{1-2m^*(r_0)/r}},
\end{equation}
which is indeed proportional to
$\gamma\left(\frac{3}{2}
,\frac{r_0^2}{4\alpha}\right)$.  Now,
$r\approx r_0$ for any small interval $I$ containing
$r_0$, so that, for all practical purposes, $\ell$
is proportional to $\gamma\left(\frac{3}{2}
,\frac{r^2}{4\alpha}\right)$ on this interval.
As a result, the smeared portion of the radius,
which is necessarily small, is proportional to
\begin{equation}\label{E:smradius}
  a_0\int^{(r-a_0)^2/4\alpha}_0
  \frac{2}{\sqrt{\pi}}\,\sqrt{t}\,e^{-t}\,dt,
  \quad r\ge a_0.
\end{equation}
(Based on the expression for $\ell$, the constant
of proportionality would not be the same for
every $a_0$.  However, in the qualitative
discussion below, the constant of proportionality
has no bearing on the outcome and can therefore
be taken as unity for any particular $a_0$.)
For this interpretation to make sense,
we have to treat $m$ as a constant, just as $a_0$
is treated as a constant in (\ref{E:mass2}).  This
is not a new assumption: even Eq. (\ref{E:mass1})
assumes, unavoidably, a fixed position at $r=0$.

One can argue that in noncommutative geometry any
measured quantity will entail a degree of uncertainty.
Since the stability question centers around the effect
of a radial perturbation on the shell, we need to
compare this effect on the two types of surfaces,
smeared and unsmeared.  To do so, the values of the
other measured quantities need not be known precisely,
as we will see in the next section.
\section{Schwarzschild wormholes}

Recall that for a Schwarzschild spacetime we have from
line element (\ref{E:line1}) that $f(r)=1-2M/r$.  So by
Eq. (\ref{E:potential})
\[
 V(a)=1-\frac{2M}{a}-4\pi^2a^2\sigma^2=1-\frac{2M}{a}-
   4\pi^2a^2\sigma_0^2\left(\frac{a_0}{a}\right)
   ^{4+4\omega},
\]
making use of Eq. (\ref{E:solution}).  From Eq.
(\ref{E:sigma}) with $\dot{a}=0$
\begin{equation*}
\sigma_0 = - \frac{1}{2 \pi a_0 }\sqrt{1-
   \frac{2M}{a_0}}.
\end{equation*}
Hence
\[
  V(a)=1-\frac{2M}{a}-\left(1-\frac{2M}{a_0}\right)
    \frac{a_0^{2+4\omega}}{a^{2+4\omega}}.
\]
The first requirement, $V(a_0)=0$ is met, but not the
second.  From
\begin{equation*}
  V'(a_0)=\frac{2M}{a_0^2}-\left(1-\frac{2M}{a_0}\right)
   a_0^{2+4\omega}(-2-4\omega)a_0^{-3-4\omega}=0,
\end{equation*}
we obtain the condition
\[
  \omega=-\frac{1}{2}\frac{a_0/M-1}{a_0/M-2}.
\]
Substituting in $V''(a)$ and simplifying, we obtain
\begin{equation}\label{E:doubleprime}
  V''(a_0)=\frac{2}{a_0^2}\frac{-1}{a_0/M-2}>0.
\end{equation}
Since $a_0/M-2$ must be greater than zero to avoid an
event horizon, the last condition cannot be met.  As
a result, there are no stable solutions for the
Schwarzschild case.

Because of its simplicity, Eq. (\ref{E:doubleprime})
provides a convenient bridge to analyzing the
smearing effect in noncommutative geometry, i.e.,
the effect of having a smeared surface.  From
(\ref{E:smradius}),
\begin{equation}\label{E:V}
  V''(r)=\frac{2}{a_0^2}\frac{-1}{(a_0/M)\int_0
   ^{(r-a_0)^2/4\alpha}\frac{2}{\sqrt{\pi}}\sqrt{t}
       e^{-t}dt-2}>0.
\end{equation}
(The reason for the change in notation is that $V''$
is now a function of $r$ in the neighborhood of
$r=a_0$.)  Condition (\ref{E:V}) can be easily met
if the smearing is substantial enough, especially
if $a_0/M$ is reasonably close to 2.

To allow a comparison to the more complicated forms
discussed later, let us consider the plot of $V''(r)$
in the neighborhood of the shell.  Even though we are
primarily interested in the qualitative features, we
need to choose some specific values for the
parameters to obtain a plot.  Suppose we arbitrarily
choose $\alpha=0.01$ and $a_0/M=5$.  (For the purpose
of illustration, $M$ is assumed to be equal to unity.)
Being arbitrary choices, the fact that these
parameters are smeared quantities is now irrelevant.
(This is also born out in the graphs, as we will see
shortly.)  For later convenience we will include
$a_0^2/2$ on the left side: being a positive
quantity, it cannot affect the sign of $V''(r)$ in
Eq. (\ref{E:V}).  So we have
\begin{equation}\label{E:Schwarz1}
  \frac{a_0^2}{2}V''_1(r)=\frac{-1}{(a_0/M)\int_0
   ^{(r-a_0)^2/4\alpha}\frac{2}{\sqrt{\pi}}\sqrt{t}
       e^{-t}dt-2}>0.
\end{equation}
The plot, shown in Fig. 1, assumes smearing in both
the inward and outward radial directions.

\begin{figure}[tbp]
\begin{center}
\includegraphics[width=0.6\textwidth]{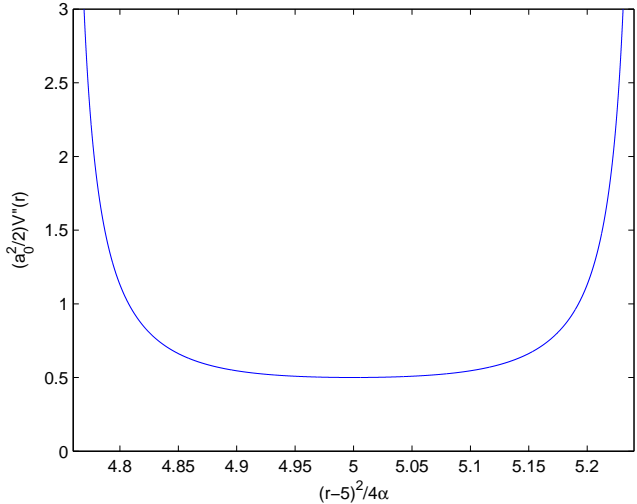}
\end{center}
\caption{The Schwarzschild wormhole: $V''>0$ near
$r=a_0$.}
\end{figure}

Since we are dealing here with a pure translation,
\begin{equation}\label{E:Schwarz2}
  \frac{a_0^2}{2}V''_2(r)=\frac{-1}{(a_0/M)\int_0
   ^{r^2/4\alpha}\frac{2}{\sqrt{\pi}}\sqrt{t}
       e^{-t}dt-2}
\end{equation}
has exactly the same shape for the same $\alpha$
(Fig. 2).  So there is no need to translate $V''(r)$ to
\begin{figure}[ptb]
\begin{center}
\vspace{0.5cm} \includegraphics[width=0.6\textwidth]{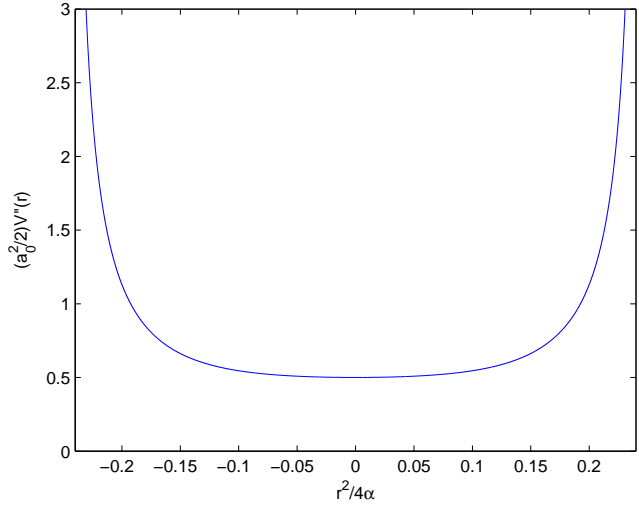}
\end{center}
\caption{The graph of Fig. 1 moved $a_0$ units to the left.
}%
\label{fig3}%
\end{figure}
determine the effect of the smearing.  The figures
show that $V''(r)$ is positive around $r=a_0$, thereby
yielding a small region of stability, i.e., a small
interval around $r=a_0$ where $V(r)$ is concave up.
As $\alpha$ gets closer to zero, the Gaussian curve,
Eq. (\ref{E:M}), is reduced in width, so that
the region of stability gets ever more narrow:
Fig. 3 shows $(a_0^2/2)V''(r)$ for $\alpha=10^{-10}$.
\begin{figure}[ptb]
\begin{center}
\vspace{0.5cm} \includegraphics[width=0.6\textwidth]{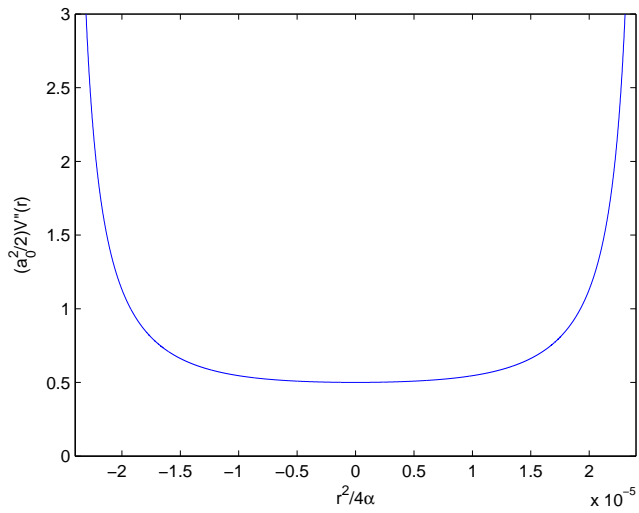}
\end{center}
\caption{The Schwarzschild wormhole with $a_0/M=5$ and
$\alpha=10^{-10}$.  The region of stability is much
reduced, implying that the wormhole is only stable to
very small radial perturbations.
}%
\label{fig3}%
\end{figure}
It is important to realize that the graph retains its
basic shape regardless of the size of $\alpha$ or the
size of $a_0/M$.

\emph{Remark 1:}  The invariance of the shape of the
graphs shows even more clearly why, qualitatively
speaking, the smearing of the parameters involved has
no bearing on the stability analysis.  Other
parameters, such as $\sigma$ and the pressure
$\mathcal{P}$ do not come into play at all at
this point, even though they are part of the
dynamic analysis of the original shell $r=a$
in the GR case.

As a final comment, as $\alpha$ gets close to
zero, the region of stability becomes vanishingly
small, and the smaller the interval of concavity for
$V$, the smaller the radial perturbations allowed.

\emph{Remark 2:} Since the smearing effect is
necessarily small, the most important applications
may very well be found in the quantum regime:
submicroscopic thin-shell wormholes with equation
of state $\mathcal{P}=\omega\sigma$, $\omega<0$,
would be stable in a Schwarzschild spacetime.

\section{Reissner-Nordstr\"{om} wormholes}

For a Reissner-Nordstr\"{om} spacetime, the starting point
is
\[
   f(r)=1-\frac{2M}{r}+\frac{Q^2}{r^2},
\]
where $M$ and $Q$ are the mass and charge, respectively,
of the black hole.  If $0<|Q|<M$, the black hole has two
event horizons at $r=M\pm \sqrt{M^2-Q^2}$ (and none if
$|Q|>M$).  Here we have
\begin{equation*}
  V(a)=1-\frac{2M}{a}+\frac{Q^2}{a^2}-\left(1-
  \frac{2M}{a_0}+\frac{Q^2}{a_0^2}\right)\left(\frac{a_0}{a}
     \right)^{2+4\omega}.
\end{equation*}
Once again, $V(a_0)=0$.  Following the same procedure
discussed in the previous section, $V'(a_0)=0$ yields
$\omega$ and $V''(a_0)$.  [See Kuhfittig (2010) for
details.]  The result is
\begin{equation}\label{E:Reissner}
  V''(a_0)=\frac{2}{a_0^2}\frac{-a_0/M-(Q^2/M^2)
  [1/(a_0/M)]+2Q^2/M^2}{(a_0/M)^2-2a_0/M+Q^2/M^2}>0.
\end{equation}
As in the Schwarzschild case, $V''$ is a function of
$a_0$; $Q/M$ is fixed.  It is also shown that for a
stable wormhole we must have
\begin{equation}\label{E:naked}
  \frac{|Q|}{M}>\frac{a_0/M}{\sqrt{2(a_0/M)-1}}.
\end{equation}
To meet this condition.\, $|Q|$ would have to exceed
$M$.  The result is a naked singularity for the black
hole.

As before, since we now have a smeared surface, we
replace $a_0/M$ by
\[
  \frac{a_0}{M}\int^{r^2/4\alpha}_0
  \frac{2}{\sqrt{\pi}}\sqrt{t}e^{-t}dt.
\]
(Recall that there is no need to translate the
curve by replacing $r$ by $r-a_0$.)  We are
primarily interested in a comparison to the
GR case.  So we retain $Q/M$ as a fixed parameter,
allowing us to concentrate on the smeared surface,
which is subject to the radial perturbation.

\emph{Remark 3:}  As discussed in the previous
section, in noncommutative geometry all measured
quantities entail a degree of uncertainty, including
$Q/M$. The precise value is not needed, however, to
draw the conclusion concerning stability, also
reiterated next.

If we now arbitrarily let $a_0/M=3$, then inequality
(\ref{E:naked}) yields $Q>1.34M$.  To show that the
wormhole has a stable region without requiring a
naked singularity, we choose $Q/M=1.2$ and $\alpha
=0.1$.  Denoting the lower incomplete gamma function
by $\gamma$, we plot
\begin{equation*}
  \frac{a_0^2}{2}V''(r)=\frac{-3\gamma-(1.2)^2\cdot
  \frac{1}{3}\gamma+2(1.2)^2}{(3\gamma)^2-2(3)
    \gamma+(1.2)^2}.
\end{equation*}
The graph is shown in Fig. 4.  This time the interval is
\begin{figure}[ptb]
\begin{center}
\vspace{0.5cm} \includegraphics[width=0.6\textwidth]{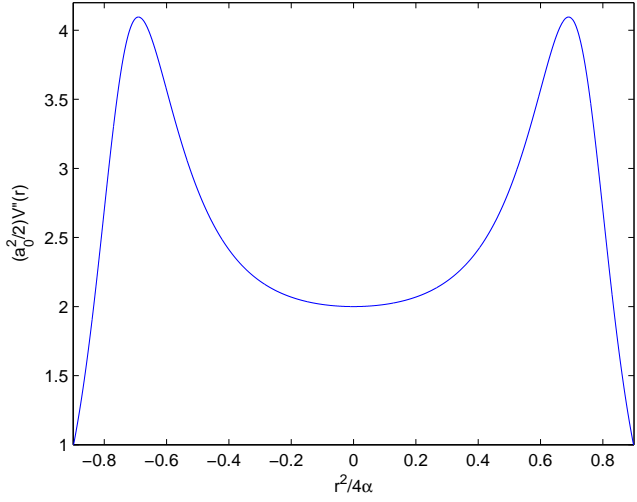}
\end{center}
\caption{The Reissner-Nordstr\"{o}m wormhole with $Q/M=1.2$
and $\alpha=0.1$, showing a region of stability without
the need for a naked singularity.
}%
\label{fig3}%
\end{figure}
made wide enough to show that $V''(r)$ eventually becomes
negative.  As in the Schwarzschild case, a smaller $\alpha$
reduces the region of stability.

As another example, Fig. 5 depicts the Reissner-Nordstr\"{o}m
\begin{figure}[ptb]
\begin{center}
\vspace{0.5cm} \includegraphics[width=0.6\textwidth]{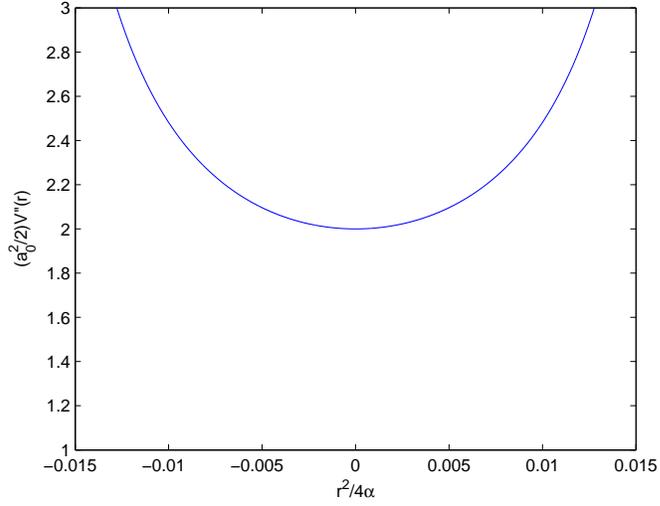}
\end{center}
\caption{The Reissner-Nordstr\"{o}m wormhole with $a_0/M=12$,
$Q/M=0.5$, and $\alpha=0.001$.
}%
\label{fig3}%
\end{figure}
case with $a_0/M=12$, $Q/M=0.5$, and $\alpha=0.001$.  As
expected, the shape has remained similar.

\section{De Sitter and anti-de Sitter wormholes}
\noindent In the presence of a cosmological constant,
$f(r)=1-2M/r-(1/3)\Lambda r^2$.  For the de Sitter case,
$\Lambda>0$.  To keep $f(r)$ from becoming negative,
$\Lambda M^2\le 1/9$.  It is shown by Kuhfittig (2010)
 that
\begin{equation}\label{E:deSitter1}
  V''(a_0)=\frac{2}{a_0^2}\frac{-1+3\Lambda M^2(a_0/M)^2
   -(2/3)\Lambda M^2(a_0/M)^3}{a_0/M-2-
   (1/3)\Lambda M^2(a_0/M)^3}>0.
\end{equation}
Here $V''$ is a function of $a_0$ with $\Lambda M^2$ fixed.
In the de Sitter case, the thin-shell wormhole is stable,
if, and only if,
\begin{equation}\label{E:deSitter2}
  1-\frac{2}{a_0/M}-\frac{1}{3}\Lambda M^2\left(
    \frac{a_0}{M}\right)^2<0.
\end{equation}
Choosing $a_0/M=5$ (arbitrarily), we obtain $\Lambda M^2
>0.07$, required for a stable solution.  To test the smearing
effect, we choose $\Lambda M^2=0.045$ (again subject to some
uncertainty).  The graph of
\begin{equation}
  \frac{a_0^2}{2}V''(r)=\frac{-1+3(0.045)(5^2)\gamma^2-
  (2/3)(0.045)(5^3)\gamma^3}{5\gamma-2-(1/3)(0.045)(5^3)\gamma^3}
\end{equation}
is shown in Fig. 6.  Once again, we see a small region of
\begin{figure}[ptb]
\begin{center}
\vspace{0.5cm} \includegraphics[width=0.6\textwidth]{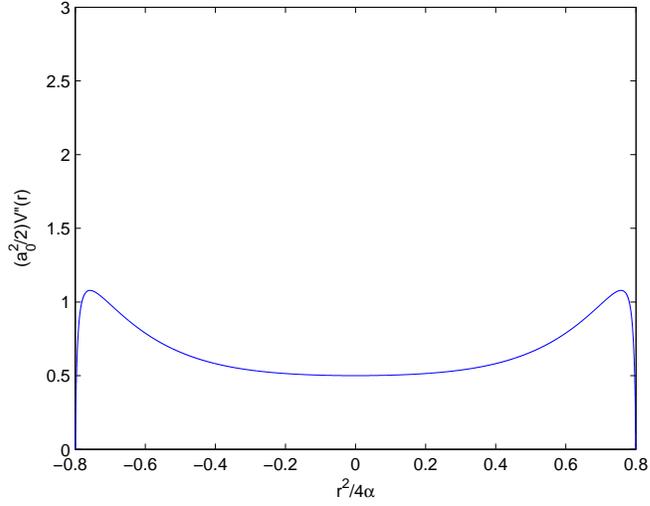}
\end{center}
\caption{The de Sitter wormhole with $\Lambda M^2=0.045$ and
$\alpha=0.1$.
}%
\label{fig3}%
\end{figure}
stability even though $\Lambda M^2$ is much less than 0.07.

In the anti-de Sitter case ($\Lambda<0$), the wormhole is
stable whenever $a_0/M>4.5$ and
\begin{equation}\label{E:anti}
   \Lambda M^2<\frac{1}{(a_0/M)[3(a_0/M)-(2/3)(a_0/M)^2]}.
\end{equation}
Choosing $a_0/M=5$ again, $\Lambda M^2<-0.12$ for a stable
solution.  If we choose $\Lambda M^2=-0.08$, thereby
violating the condition, the smearing effect produces a
plot for $(a_0^2/2)V''(r)$ that is similar to the graph
in Fig. 6.

In summary, while the wormholes in the de Sitter and
anti-de Sitter spacetimes normally require sufficiently
large $|\Lambda|$ to get a stable solution, a
noncommutative geometry background allows much smaller
values of $|\Lambda|$.

\section{Conclusion}
\noindent This paper reexamines a special class of
thin-shell wormholes known to be unstable to linearized
radial perturbations in classical general relativity
(GR).  In the framework of noncommutative geometry,
however, small regions of stability are obtained,
thereby allowing small radial perturbations.  The
size of the stability region depends on the parameter
$\alpha$, which is used to measure the degree of
smearing due to the intrinsic uncertainty.

For the four spacetimes considered, regions of
stability were obtained (1) for the normally unstable
Schwarzschild wormhole, (2) for the
Reissner-Nordstr\"{o}m wormhole without requiring a
naked singularity, and (3) for both the de Sitter
and anti-de Sitter wormholes for $|\Lambda|$ much
smaller than the values required in a GR setting.

\begin{center}
\bf{Acknowledgment}
\end{center}
The author would like to thank Vance Gladney for helpful
discussions.

\end{document}